\documentclass[runningheads]{llncs}

\usepackage[utf8]{inputenc}
\usepackage{hyperref}
\usepackage{xspace,url,graphicx,soul,array,siunitx}
\usepackage{pgfplots}\pgfplotsset{compat=1.14}
\usepackage{tikz-uml}

\newcommand{\Year}{\the\year\xspace}
\newcommand{\mapc}{\textsf{Multi-Agent Programming Contest}\xspace}
\newcommand{\mas}{\textsf{MAS}\xspace}
\newcommand{\shortmapc}{\textsf{MAPC}\xspace}
\newcommand{\contest}{\textsf{Contest}\xspace}
\newcommand{\massim}{\textsl{MASSim}\xspace}
\newcommand{\EIS}{\textsf{EIS}\xspace}
\newcommand{\EISMASSim}{\textsf{EISMASSim}\xspace}

\newcommand{\smartja}{\emph{SMART-JaCaMo}\xspace}
\newcommand{\flisvos}{\emph{Flisvos}\xspace}

\newcommand{\busyb}{\emph{BusyBeaver}\xspace}
\newcommand{\pucrs}{\emph{PUCRS}\xspace}

\newcommand{\psj}{\emph{PSJ}\xspace}

\newcommand{\fitbut}{\emph{FIT BUT}\xspace}
\newcommand{\dtu}{\emph{GOAL-DTU}\xspace}
\newcommand{\lfc}{\emph{LFC}\xspace}
\newcommand{\trg}{\emph{TRG}\xspace}

\soulregister{\pucrs}{0}
\soulregister{\flisvos}{0}
\soulregister{\busyb}{0}
\soulregister{\smartja}{0}
\soulregister{\psj}{0}

\usepackage[color=yellow]{todonotes}

\begin{document}

\title{The Multi-Agent Programming Contest:\\ A résumé}
\subtitle{Comparing agent systems 2005-2019}
\titlerunning{The MAPC: A résumé}

\author{Tobias~Ahlbrecht\inst{1}\orcidID{0000-0002-4652-901X} \and
Jürgen~Dix\inst{1}\orcidID{0000-0002-8528-1440} \and
Niklas~Fiekas\inst{1}\orcidID{0000-0002-8369-4890} \and
Tabajara~Krausburg\inst{1,2}\orcidID{0000-0002-8252-4099}}
\authorrunning{T. Ahlbrecht et al.}

\institute{Department of Informatics, Clausthal University of Technology, Clausthal-Zellerfeld, Germany \\ \email{\{tobias.ahlbrecht,dix,niklas.fiekas\}@tu-clausthal.de} \and
School of Technology, Pontifical Catholic University of Rio Grande do Sul, Porto Alegre, Brazil\\
\email{tabajara.rodrigues@edu.pucrs.br}}

\maketitle

\begin{abstract}
    The Multi-Agent Programming Contest, \shortmapc, is an annual event organized since 2005 out of Clausthal University of Technology. Its aim is to investigate the potential of using decentralized, autonomously acting intelligent agents, by providing a complex scenario to be solved in a competitive environment. For this we need suitable benchmarks where agent-based systems can shine. We present previous editions of the contest and also its current scenario and results from its use in the 2019 \shortmapc with a special focus on its suitability. We conclude with lessons learned over the years.

    \keywords{multi-agent systems  \and decentralized computing \and cooperation \and artificial intelligence \and simulation platforms.}
\end{abstract}

\section{Introduction}
The original aim of our contest, back in the humble beginnings in 2005, was to provide a platform for comparing and evaluating systems based on \emph{computational logic}, mainly developed for knowledge representation purposes.

We wanted to develop an interesting yet  simple, but non-trivial, scenario for testing  systems based on different paradigms. At that time, many knowledge-based approaches  were developed as  smallish  PhD projects: A prototype was implemented but never seriously compared against  other such systems.

Emphasis was  put on the \emph{evaluation} and \emph{comparison} of systems, not on finding an optimal solution of a particular scenario. The creation of a scenario was always driven 
by the need to determine  the features that a system should possess for successfully solving a complicated task. We never wanted to honor a smart idea for a solution, but the features and technology that help to tackle the problem at hand.

\subsection{Agents}

During the years, the systems we compared turned more and more into those based on \emph{agent programming languages} \cite{jaamas} or genuine multi-agent systems (\mas) implemented  in classical programming languages. The scenarios
became more complex with an increasing number of agents needed to solve the task.

In contrast to many other contests, several decisions have been taken a priori:
\begin{itemize}
    \item not to impose any restrictions on the software used;
    \item not to find or compare \emph{tricky algorithms} to solve the scenario, rather we wanted to evaluate the capabilities of the system to express and model suitable constructs for dealing with the scenario;
    \item not to consider the perfect implementation or high performance of a system;  in particular, we never considered  real-time aspects, which are important for e.g.~computer games.  
\end{itemize}

The last bullet above reflects the situation in agent programming for many years (still today, but to a lesser extent): Agent languages are still not on par with classical programming languages in terms of their efficient implementation and their maturity concerning software engineering aspects. We therefore decided to refrain from this particular aspect.

In our \shortmapc, each participating team develops a group of agents (during the 5-7 months between the 
announcement of the scenario and the contest),  which remotely connect to our \shortmapc server where the scenario is being run. The \shortmapc server sends the current game state in the form of percepts to each agent and expects an executable action in return. The gathered actions are executed and the game state is advanced. This cycle is repeated until a predefined number of steps is reached.
The remote nature of the contest also keeps the responsibility of running the agents with the participants.

The available time for each simulation step must include the latency of the internet and is, intentionally, chosen to be quite high
(4 seconds): We do not consider high performance nor real-time constraints.

In addition, we have no control about the communication within a team (shared memory or not, decentralized or not). Consequently, we could not directly enforce decentralized approaches---only by designing the scenario in a way that favors them.

We also never excluded classical (i.e. non-agent) programming languages and frameworks from being used. In fact, we almost always had non-agent entries take part in our contest and some performed very well. Obviously, one can use a classical programming language and implement certain agent-techniques that are suitable for the scenario (or use agent technology without leveraging its potential, in effect using it like a conventional programming approach).

\subsection{Goals and Purpose of the \shortmapc}

The purpose of the contest is twofold: (1)  to  find out for which applications agent-oriented features pay off, as opposed to features available in classical programming languages, and 
(2) to compare and test the versatility and suitability of agent languages or platforms. The main questions guiding us for the last decade were: \emph{To which extent, if any, are agent-oriented programming languages superior to classical languages and for which problem instances do we gain the most?}

The difference between agents and classical, more centralized paradigms is, to a great deal, autonomy, communication, cooperation and to strike a good balance between proactiveness and reactiveness.

Clearly, any (new) feature can be implemented in any (Turing-complete) programming language, but one would hope an agent language to be more versatile and efficient or offering built-in features for elegantly programming a solution.

Ultimately we do not want to compare problem solutions, but we want to compare agent languages among themselves and against classical programming languages.

Therefore we always try to develop our scenarios in such a way, that no smart solution will be sufficient, but the \emph{interplay of various acting entities and their emerging features}.

In the end, we are especially interested in which technologies the teams used and to which degree, how difficult (or easy) it was to use agent-based options and which aspects where especially straightforward or challenging to design and implement. 

The contest is an attempt to shed some light on these questions: When and to what extent do agent-oriented features pay off? Is there a particular complexity of the problem that makes these approaches beneficial? Or not at all?

Moreover, we aim to support educational efforts in the design and implementation of agent systems by providing each year a ready, off-the-shelf package:  this is specially tailored for the classroom and can be used in a course on agent systems of any level.
We noticed in our experience that the competition idea is especially attractive for students and results in a very engaging work atmosphere.

\subsection{Related work}
Many similar competitions have been and are still being held, while most do not explicitly focus on multi-agent systems. We discuss some of them that are  related to \shortmapc and are still active nowadays.

Directly involving agents, the \emph{(Power) Trading Agent Competition}\footnote{\url{www.powertac.org}}~\cite{ketter2013power} provides a trading-related scenario in the energy market. However, each team only consists of a single ``broker'' agent, requiring no cooperation or coordination. The goal here is to see how agents can autonomously solve supply-chain problems.

Probably the best-known are the various \emph{RoboCup Simulation Leagues}\footnote{\url{www.robocup.org}}. RoboCup ranges over a variety of different domains like soccer, disaster response, and industrial logistics. Each league focuses on a specific problem that must be addressed by competitors. For instance, in RoboCupRescue two major leagues are organized on: (i) robots; and (ii) agent simulation. The first centers around (virtual) robots and less around abstract agents. For example, agents have noisy virtual sensors or may be subjected to complex physics, focusing on realism. The agent simulation league provides virtual agents placed on a map of a city that has been damaged by an earthquake event. Competitors focus on different self-isolated AI problems (e.g., task allocation) provided by the contest developing algorithms for them~\cite{visser2015robocuprescue}. In addition, all teams have to give a presentation on their solution, which counts towards their final score.

There is also a number of challenges targeting specific problem domains, e.g. the \emph{International Planning Competition}~\cite{vallati20152014}. Here, of course planning is in the limelight, while in our contest it is only one possible component of an agent team.
At the other end, the \emph{General Game Playing}~\cite{genesereth2005general} competitions do not focus on one particular feature but on the ability of general AI systems to play an arbitrary game upon receiving its rules.

Finally, there are more than a few challenges focusing on finding (autonomous) solutions for existing commercial games, like the \emph{Mario AI Championship}\footnote{\url{www.marioai.org}}~\cite{mario} or the \emph{Student StarCraft AI tournament}\footnote{\url{www.sscaitournament.com}}, or specifically designed games like \emph{BattleCode}\footnote{\url{www.battlecode.org}}. The goal here is usually to benchmark game AI techniques and algorithms.

We would also like to mention a new challenge, the Intention Progression Competition\footnote{\url{www.intentionprogression.org}}, which focuses on a specific issue within agent systems: The Intention Progression Problem, i.e. the decision of agents about how to proceed with their given intentions and plans in order to reach their goals. Thus, solutions for the \shortmapc could be seen as specific challenges in the IPC, while solutions for the IPC could be used in agent platforms that participate in the \shortmapc.

\subsection{Structure of this work}
We started by introducing the \mapc. In Section \ref{sec:framework}, we introduce the simulation platform, followed by the history of the \contest in Section \ref{sec:history}. Afterwards, we present the newest scenario and the results of the \shortmapc 2019 in Section \ref{sec:2019}. We conclude with lessons learned during the contest in general and in the latest installment in particular. Throughout the article, we focus especially on the scenario aspect of the \shortmapc. 

\section{The MASSim framework}\label{sec:framework}

The first edition of the \shortmapc  in 2005 presented a simple scenario that had to be
implemented in its totality by each participant and delivered as an executable.

In 2006, the \massim infrastructure was introduced: an extensible simulation server that
provides the environment facilities. Agent programs can connect through the network to a \massim server while  agents  run in the competitors' own computer infrastructure. Since then, the format of the \shortmapc  has been that of two teams competing against each other for performance
in each simulation, and the overall winner of the contest defined by summing up the points after all participants
have competed in simulations against each other, in a regular sports tournament fashion.

All simulations are run in a step-by-step manner. In each step all agents execute their actions simultaneously
from the point of view of the server, and there is a time limit within which agents must choose an action
(otherwise they are regarded as a \texttt{no-op}). In the beginning of each step's cycle, the server sends each
agent its current percepts of the environment, and waits for the response that specifies the action to execute.

When the responses from all agents are received or when the timeout limit is reached, all received actions are
executed in \massim, and the agents' percepts for the next step are calculated. This cycle is repeated for a fixed
number of steps, and then a winner is decided according to scenario-specific criteria.

\massim is fully implemented in Java, and the information exchange with the agent programs is made through JSON (formerly XML)
messages.

In early 2017, \massim was completely rewritten. We switched from having both a Java RMI based monitor and a web monitor to a single web-based monitor. Also, we abandoned the former plug-in architecture in favor of a yearly package, which helped in keeping the package small and freed us of having to keep \massim backwards-compatible to all previous scenarios.\footnote{You can still play the old releases using their respective packages.} This rewrite also allowed us to create a platform with more than two concurrent teams in mind. While we have not used this yet, it remains a tempting option for future sce\-na\-ri\-os.

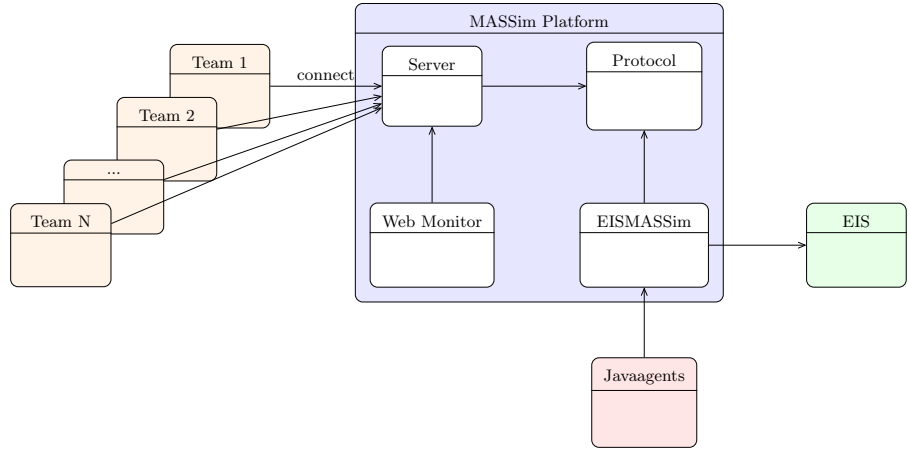
\begin{figure}[ht]
\centering
\resizebox{0.99\textwidth}{!}{
\begin{tikzpicture}
    \begin{umlstate}[name=massim, fill=blue!10]{MASSim Platform}
        \begin{umlstate}[y=-3, x=4, name=eismassim, fill=white]{EISMASSim}
        \end{umlstate}
        \begin{umlstate}[name=server, fill=white]{Server}
        \end{umlstate}
        \begin{umlstate}[x=4, name=protocol, fill=white]{Protocol}
        \end{umlstate}
        \begin{umlstate}[y=-3, name=monitor, fill=white]{Web Monitor}
        \end{umlstate}
    \end{umlstate}
    \begin{umlstate}[x=8, y=-3, name=eis, fill=green!10]{EIS}
    \end{umlstate}
    \begin{umlstate}[x=4, y=-6, name=agents, fill=red!10]{Javaagents}
    \end{umlstate}
    \umltrans{eismassim}{eis}
    \umltrans{server}{protocol}
    \umltrans{eismassim}{protocol}
    \umltrans{monitor}{server}
    \umltrans{agents}{eismassim}
    
    \begin{umlstate}[x=-4, name=team1, fill=orange!10]{Team 1}
    \end{umlstate}
    \umltrans[arg={connect},pos=0.5]{team1}{server}
    
    \begin{umlstate}[x=-5, y=-1, name=team2, fill=orange!10]{Team 2}
    \end{umlstate}
    \umltrans{team2}{server}
    
    \begin{umlstate}[x=-6, y=-2, name=team3, fill=orange!10]{...}
    \end{umlstate}
    \umltrans{team3}{server}
    
    \begin{umlstate}[x=-7, y=-3, name=teamn, fill=orange!10]{Team N}
    \end{umlstate}
    \umltrans{teamn}{server}
\end{tikzpicture}
}
\caption{The \massim infrastructure.}
\label{FIG:massim}
\end{figure}

Figure~\ref{FIG:massim} displays the current architecture. The server package is responsible for running the simulations and handling connections to all agents.
The protocol has been extracted into a self-contained package that can be used to create compliant messages. It helps both with parsing JSON data into Java objects and transforming Java message objects into their JSON representation. Thus, it is e.g. used by the server to create messages for all agents. The protocol is also used by the \EISMASSim component, which can be used by agent platforms to connect to the server. This component translates perception and action messages into actions and percepts according to the \EIS (Environment Interface Standard~\cite{behrens2011towards}) and vice versa. We also provide a sample implementation of agents using \EISMASSim in the \texttt{Javaagents} package. Participants using Java-based platforms may connect to the server by integrating \EISMASSim, using the protocol package, or, just as non-Java-teams, parse and build their own JSON messages according to the protocol.

The current \massim package is fully open-source and openly available (\url{https://multi-agentcontest.org/2019}).
It is not only used for the \shortmapc, but has also proved useful both for researchers testing their advancements in the field, and in the classroom, aiding the teaching of the multi-agent programming paradigm (\url{https://multi-agentcontest.org/massim-in-teaching}).

\section{History and evolution of the contest}\label{sec:history}

We can roughly divide the contest into two phases. 
In the early phase, there was not much cooperation among the agents: They acted more or less on their own. 
This led us to reconsider our scenario and we ended up with the \emph{Agents on Mars} scenario, where we experienced some really interesting games. 
This then evolved into the \emph{Agents in the City} (or simply \emph{City}) scenario, which was even more realistic as it considered agents acting in a real city using actual city maps. 
We then adapted the \emph{City} scenario, removing some of its complexity (regarding implementation effort for the participants) and incorporating features we think were interesting from previous scenarios, which led to the \emph{Agents Assemble} scenario, which we will present and analyze in detail.

\subsection{Early phase}

The scenario used for the first edition of the \shortmapc (2005) consisted in a simple grid
in which agents could move to empty adjacent spaces. Food units would appear randomly through the simulation, and the objective was to collect these units and carry them to a storage location. 

\begin{figure}[ht]
		\centering
		\includegraphics[width=0.45\textwidth]{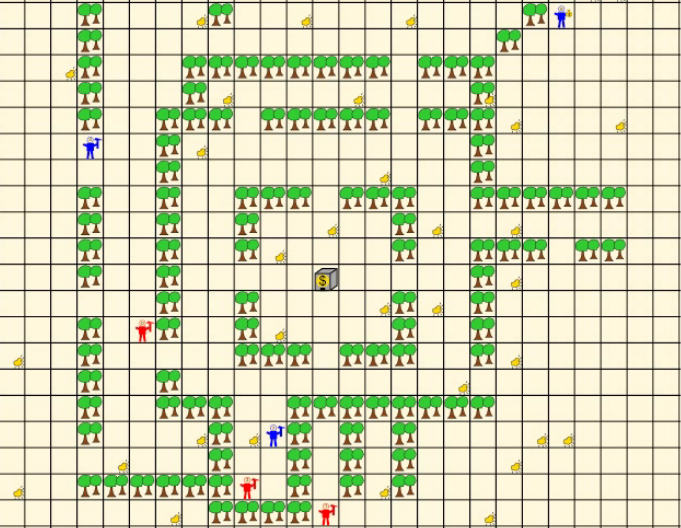}
		\vskip-0.3cm\caption{The Gold Miners scenario.}
		\label{FIG:gold-miners}
\end{figure}

The idea was refined for the second edition: \emph{Gold Miners}. 
Now the agents were to collect gold in a competitive environment against another team, and some obstacles were introduced to the grid to add some navigation complexity. 
This scenario, which was also used in the third edition of the contest, was still very simplistic, and in the proposed solutions agents acted independently of their teammates: No cooperation or coordinated behavior took place.

\begin{figure}[ht]
	\centering
	\includegraphics[width=0.45\textwidth]{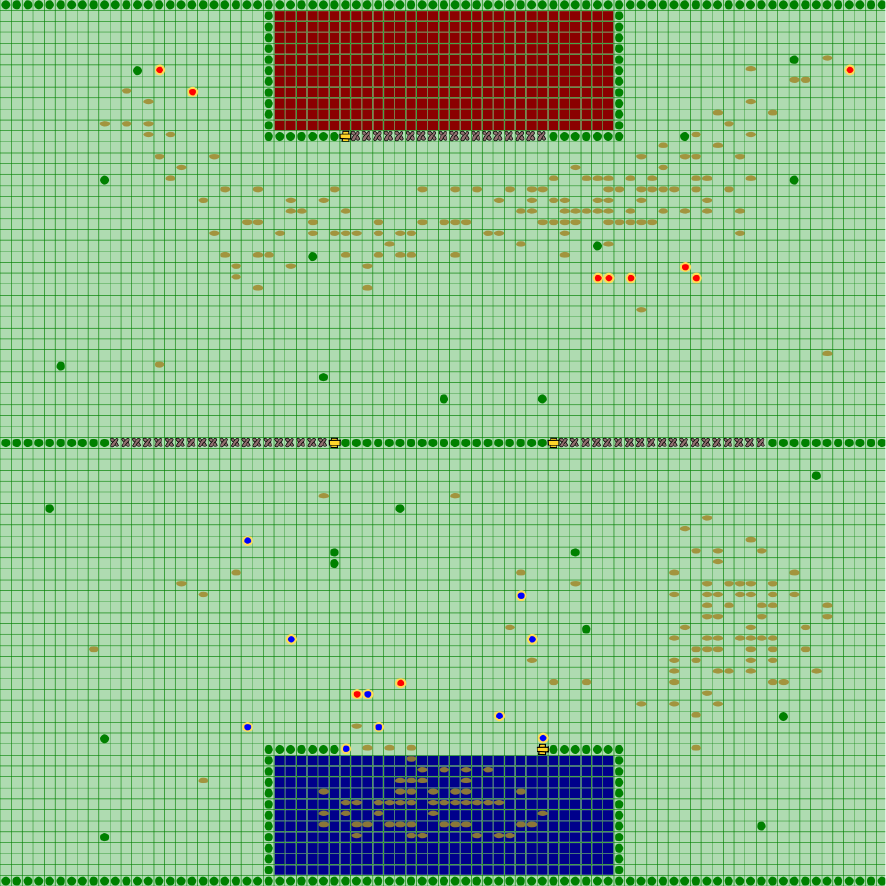}
	\vskip-0.3cm\caption{The Cows and Cowboys scenario.}
	\label{FIG:cow-herders}
\end{figure}

For the 2008-2010 editions, a new scenario was designed that \emph{demands} coordination from agents: \emph{Cows and Cowboys}. Still using a grid as the underlying map, the goal for this scenario was to lead a group of cows to a particular area of the map, the team's own ``corral'', while preventing the opponent team from doing the same. 
The cows were animated entities that reacted to the agents' positions by trying to avoid them. Solving the map required agents to coordinate their positions in order to lead big groups of cows into the corrals, whereas a single agent would in most cases disperse the group of cows and fail to lead them in the desired direction.

Even in this clearly cooperative scenario, one team found a way of letting each agent work independently, always pushing a single cow. This team promptly won the contest (though out-of-competition) and we learned that \emph{features we want to see need to be enforced rather than rewarded}, since participating teams always tend to find (and go for it) the path of least resistance. 
Thus, a flocking algorithm for cows was introduced, which made the cows form groups and avoid agents more strongly. This allowed good teams to capture entire herds with the right agent formations, while single agents could not achieve anything anymore. 
In addition, fences were added as another cooperative element: Agents had to stand on switches to open them and communicate to get all agents and cows safely through.
In that way we achieved some cooperation among agents and saw the first interesting games.

\subsection{Agents on Mars}

The \emph{Agents on Mars} scenario was used from 2011--2014.
It turned out to be an important step in the contest's evolution, as it introduced many innovative features and increased the game's complexity.
The map took the form of a weighted graph representing the surface of the planet Mars (we always based the scenario on a fictitious story). 
The agents represent \emph{All Terrain Vehicles} of different kinds, and their goal in the game is to discover the best water wells by exploring the map and then to keep control of as many  wells as possible. This was done  by placing themselves in specific formations that ensure a covering of an area containing the wells while keeping rival agents aside. 

\begin{figure}[ht]
	\centering
	\includegraphics[width=0.45\textwidth]{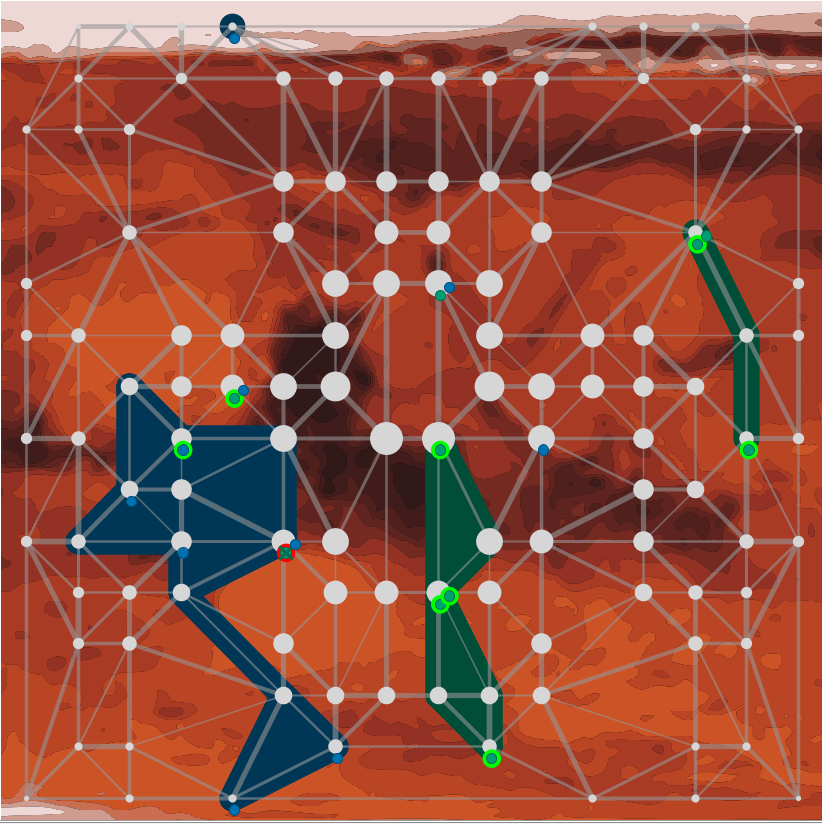}
	\vskip-0.3cm\caption{The ``Agents on Mars'' scenario.}
	\label{FIG:agents-on-mars}
\end{figure}

These agents were much more complex entities than in the previous scenarios: They had a rich set of actions to choose from, in contrast to just moving around the map. Furthermore, they dealt with a set of internal parameters that could vary through the simulation---\emph{Energy}, \emph{Visibility Range}, \emph{Health} and \emph{Strength}.

The evolution in the complexity of the scenario has remained on par with the evolution of multi-agent programming technologies used by the participating teams. 
A good quality of the teams has been reached and resulted in interesting games. Unlike previous scenarios, a (simple) strategy that works against each and every rival has
not been discovered.

\subsection{The City scenario}

\begin{figure}[ht]
  \centering
  \includegraphics[width=0.45\textwidth]{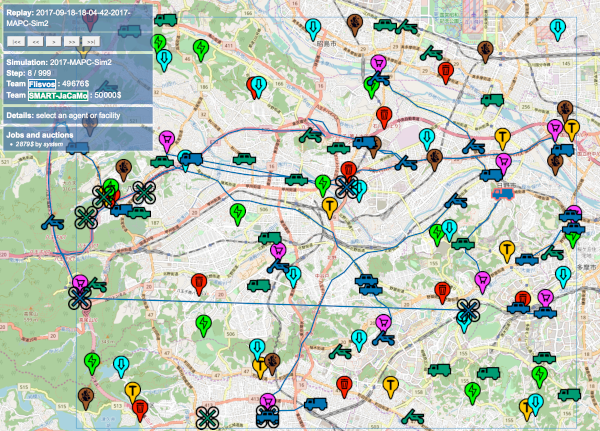}
  \vskip-0.3cm
  \caption{The ``Agents in the City'' scenario.}
  \label{FIG:city-scenario}
\end{figure}

Our previous scenario, pictured in Figure \ref{FIG:city-scenario}, was first used in 2016 and improved two times for the editions of 2017 and 2018. We started with two teams of 16 agents each moving through the streets of a digital city backed by realistic street graph data from \emph{OpenStreetMap}\footnote{\url{https://www.openstreetmap.org}}. The number of agents was then increased to 28 and 34 per team respectively.

Each team's goal was simply to earn as much money as possible by completing randomly generated jobs. These jobs required the agents to move around the city, buy certain kinds of items, cooperatively assemble these items to get new item types and finally deliver the finished products to a predefined target location. Most of these jobs were available for both agent teams simultaneously and rewarded on a \emph{first come, first served} basis, allowing for more direct competition.

Each agent had one of four distinct roles, which characterized its movement type (air- or road-bound) and speed, as well as its maximum battery and carrying capacity. 
As is tradition, the number of agents was increased for each scenario to provide a greater challenge of coordination and require some more computational effort.
Different agent roles were first introduced with the \emph{Agents-on-Mars} scenario. The roles differed by certain key attributes as well as by which action was usable by which agent.

Compared to our previous scenarios, this one required more coordination and planning among agents of the same team. Jobs may not be profitable or less profitable than co-occurring jobs. Once agents are able to identify good jobs, the real challenge is the coordination of which agent secures which items from where in order to strike a good balance between time efficiency and money spent.

For the third instance of the scenario, we added a new \emph{well} facility that teams could build and opposing teams could dismantle. To build wells, some funds had to be spent which could again be acquired by completing jobs. The wells would then generate points for as long as they existed. This change was intended to increase interaction between the teams and make the agents' actions more visible to human observers.

\subsubsection{Lessons learned in the City}

The first run in 2016 has shown once again that participants have to be coerced into using specific features of the scenario: For example, we had to make cooperative assembly mandatory in 2017.

For the second run in 2017, we noticed a problem with the many parameters controlling the random generation of simulation instances. Finding \emph{good} sets of parameters was not an easy task and required considerable testing.
Also, for the first time we experienced that a scenario should allow for a simple naive (but far from being optimal) solution to be quickly producible. This scenario instead required considerable work until first results can be seen.

Another downside was that the visualization did not (or could not) show everything that was going on in an easily discernible way. For example, it is very impractical to display for all agents which items they are currently transporting. To amend this a little, the wells were added in 2018 to have an element that plainly shows how well a team is doing aside from the current money value.

Also, interaction between the teams was very limited and only indirectly given through the availability of shared resources (i.e. items in the shops) and the competition to get a job done first in order to receive the reward. The wells were also added to have a new entity that agents of both teams could and needed to interact with.

\section{2019: Agents assemble}\label{sec:2019}
After having played the \emph{City} scenario for three consecutive years, it was once again time to come up with a fresh scenario and apply the lessons learned. We wanted to address some of the issues with the previous scenario, like visibility of agent behavior, while keeping many of the factors that made it interesting.

\subsection{Scenario}

\begin{figure}
    \centering
    \includegraphics[scale=.2]{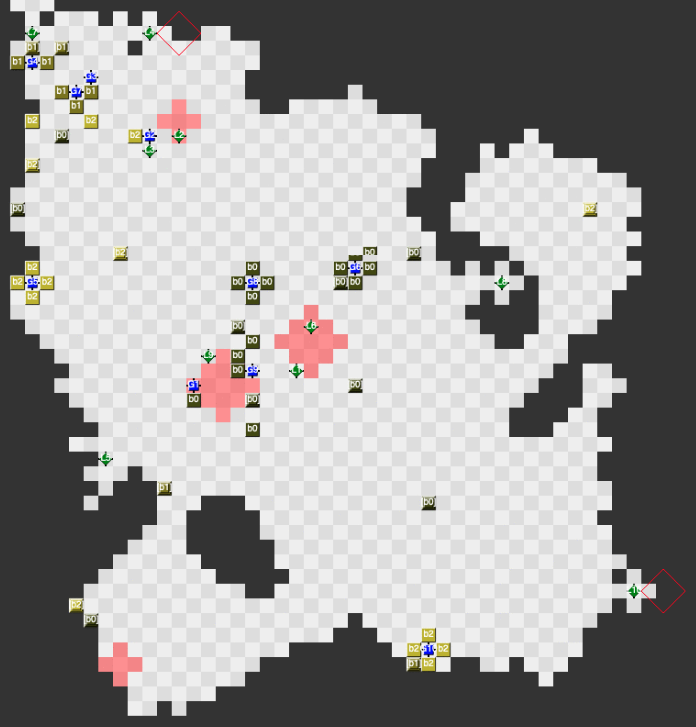}
    \caption{\shortmapc 2019 environment. Agents possess a local view of it and are required to assemble complex shapes to be delivered.}
    \label{fig:mapc_2019}
\end{figure}

In the new \emph{Agents Assemble} scenario, as the name suggest, agents again have to construct complex structures from base objects. We switched from the map- (or graph-)based environment back to a ``simple'' grid structure with obstacles (see Figure~\ref{fig:mapc_2019}), comparable to the \emph{Cow} scenario. 
The agents have to explore the grid to find \texttt{blocks} which also occupy one cell of the grid. Each agent has four ``arms'', one to each side, which can be used to pick up or connect to blocks. Blocks which are connected to an agent move in the same direction as the agent. Two adjacent blocks can also be connected to each other by two agents from the same team, when each agent is holding one of the blocks.

\begin{figure}
    \centering
    \includegraphics[scale=.3]{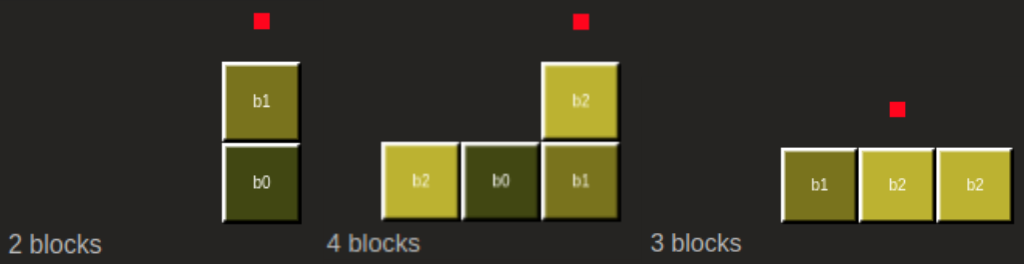}
    \caption{Some examples of tasks in which the delivery agent should be carrying the blocks at the red dot position depicted in the figure.}
    \label{fig:mapc_2019:tasks}
\end{figure}

The system then randomly creates \texttt{tasks}, which the agents have to complete to earn reward points. The team with the most points at the end of the simulation will be the winner. Each task basically describes a structure or formation of blocks that the agents have to create. We depict an example of tasks in Figure~\ref{fig:mapc_2019:tasks}. Once the shape is assembled, the agents can deliver it to one of the goal zones to receive the points.

\subsubsection{Actions}
The agents have different actions for moving around in the grid. They can move one cell in each of the four main directions per step or rotate 90 degrees. This rotation might be handy, if the agents have blocks attached. Further, there is an action to retrieve blocks from \texttt{dispensers}, which are placed in random locations and provide one specific type of block. To work with blocks, the agents have actions for attaching and detaching things to their sides and as mentioned before, two agents can use the \texttt{connect} action to join two blocks together. An agent can also break this connection between two blocks, if the blocks are attached to the agent (directly or indirectly).

To interact with the environment and other agents, the \texttt{clear} action was added. It targets a single cell and has to be ``charged'', i.e. executed a certain number of times for the same target cell before it has an effect. Once it resolves, if the cell contained an obstacle or block, these will vanish and leave an empty cell. If instead an agent occupied that cell, it will be disabled. In that case, this agent will not be able to execute actions for a certain number of steps and also, all of its attached blocks (if any) will not be attached to the agent anymore. To give each agent a chance to avoid this, the target cells to be cleared have a perceivable marker after each \texttt{clear} action, i.e. also while charging.

As always, each action has a number of specific failure codes, indicating the reason why the action could not be executed.

\subsubsection{Perception}
One of the novelties of this scenario is that agents only perceive relative coordinates. That is, at the beginning the agents cannot know where they are. Due to their limited vision range of five cells in each direction, they do not even know where they are relative to each other and have to find their teammates first.

This might favor solutions, where a local agent perspective is taken, rather than centralized approaches.

\subsubsection{Dynamic environment}
To give the agents an even greater challenge, the environment dynamically changes during the matches. This makes is harder for the agents to remember if they have already been at a place and requires more adaptability.

During each game, a number of \texttt{clear events} will occur. These work almost exactly like the \texttt{clear} action, only they affect a bigger region of the grid and after each event, new obstacles will appear randomly distributed around the center of the event.

\subsubsection{Blocks and visibility}
One drawback of the \emph{City} scenario was that it was not very interesting to watch, because most of the action did not happen in the environment. When agents bought items, these just went to their inventories. The current possessions of an agent could be displayed in a list, but it was rather difficult to keep track of multiple agents at once, not to mention all of them. Thus, in the new scenario, items (i.e. \texttt{blocks}) have received a more tangible representation, taking up considerable space in the environment. This leads to more interaction between agents and items and all of it is easily observable by human bystanders. What's more, carrying assembled shapes around becomes even more of a challenge, as the number of available routes possibly decreases.

\subsection{Participants}
This year, we had four teams participating in the \contest.

\begin{description}
    \item[FIT BUT] The team from Czech Brno University of Technology consists of three people and participated in the \contest for the first time. The agents are implemented in plain Java.
    \item[GOAL-DTU] The team from Technical University of Denmark has already participated in the \shortmapc in one form or another for many, many years and has never missed a \contest since. As the name suggests, the agents were implemented using the GOAL~\cite{goal} agent language.
    \item[LFC] The team LFC, from University of Liverpool, used JaCaMo~\cite{boissier2013multi} to implement its agent team. An additional fast downward planning component was developed to support the agents.
    \item[TRG] The single-person team TRG from the Canadian Carleton University also participated in the \contest for the first time. The agents were implemented with the Jason~\cite{jason} framework.
\end{description}

An overview of the teams is listed in Table~\ref{tab:teams}. As we can see, this year, all approaches involve Java at some level. The teams are of similar size, except for \trg. Notably though, the single-person team has invested the most time. Of the four teams, \trg and \fitbut are completely new to the \contest, while some members of \dtu and \lfc had already participated before. We also note that the GOAL solution is particularly small in terms of LOC, while the Jason-based solution is a bit larger than the average.\footnote{This does not necessarily tell us anything about GOAL or Jason though.}

\begin{table}[htbp]
    \centering
    \setlength\extrarowheight{5pt}
    \begin{tabular}{| c | c | c | c | c |}
      \hline
      & \fitbut & \dtu & \lfc & \trg \\ 
      \hline\hline
      System & Java & GOAL & JaCaMo & Jason\\
      Team size & 3 & 3 & 3 & 1\\
      Time invested (in h) & 300-400 & 200 & 200 & 500-600 \\
      Previously participated & No & Partly & Partly & No\\
      LOC & 5500-6300 & 1000 & 6800 & 9700\\
      Started & 29.8. & August & May/September & May/mid-July\\
      \hline
    \end{tabular}
    \caption{\label{tab:teams}Team overview}
\end{table}

\subsection{Tournament and results}

In the final tournament, each team plays one match against each other team, where one match consists of three simulations with different parameters. Thus, with four teams, each team had to play 9 games. Winning a simulation is awarded with three tournament points, while a draw means one point for each team. The best result a team can achieve is 27 points.

We had the teams play simulations with three different sets of parameters, so that they were less likely to optimize their systems to one particular setting. Each simulation ran for 500 steps and 10 agents per team. In the second simulation, more complex tasks, with up to 5 required blocks instead of 3, were offered. In the third simulation, we increased the chance of a random \texttt{clear event} happening from 4\% to 8\%, leading to a more uncertain environment.

The \contest was won by the JaCaMo-based solution from Liverpool's \lfc, with only one loss against \dtu and one draw against \trg out of 9 games, resulting in 22 tournament points. Runner-up is \fitbut with 15 points, while \dtu achieved 10 and \trg 5 points. We note that each team won at least one simulation, and never only because the other team failed completely. All teams presented a workable solution. 

\subsubsection{Strategies}
No team found a \emph{strategic} advantage over the others. That is, we did not see a particular strategy being used to great effect. While the agent teams approached the problem in different ways, none of these were clearly superior to all others.

The \contest winner, \lfc, implemented a strategy, where one agent was always waiting in a goal zone for its team members to deliver exactly the blocks needed for a particular task. We saw each agent always carry at most one block at a time. The shape required for the task was always assembled together with the agent waiting in the goal zone, who then submitted the task upon its completion. One advantage of \lfc was clearly the capability to ``dig'' straight lines through obstacles with repeated \texttt{clear} actions. This technique was also used by the agents at the start of each simulation, probably to find the actual boundaries of the grid environment (which was always surrounded by a wall of obstacles). \lfc implemented dynamic roles, where agents would start as explorers and later switch to specializations, e.g. assembling agents waiting in the goal zones.

\fitbut in contrast had their agents meet somewhere on their routes to connect their blocks. Thus we saw \fitbut agents walking around with complex shapes attached, which also worked very well.

\dtu agents could always be recognized by them proactively requesting as much as four blocks at a time and subsequently moving with four blocks of one type attached. While this ensured that they always had enough blocks at their disposal, it made it more difficult to navigate the map, especially during the late game when \texttt{clear events} could have already created narrow paths.

\trg alone tried a hybrid strategy. While some agents were coordinating to complete tasks, the other agents were trying to ``defend'' each goal zone by using \texttt{clear} actions on approaching opponent agents. This was an interesting decision, which unfortunately did not pay off so well, as the agents from the other teams were mostly able to circumvent these interventions. These roles were also statically assigned and did not depend on the current situation.

While we always try to build and configure the scenario in such a way, that no single best (and maybe even simple) strategy exists, we can never be sure that we succeed in this. Generally, the more features and interactions among them a scenario has, the harder it becomes to balance all of these features ``correctly'', so that no single feature can be used in an unforeseen way. Thus, in the new scenario, the rules governing the simulations were kept as simple as possible.

\subsection{Interesting simulations}
In this section, we want to take a look at some interesting simulations\footnote{Not to say that some of the simulations were not interesting!} to see how the teams compare to each other under similar circumstances.

\subsubsection{2nd simulation of \dtu vs. \lfc}
\begin{figure}[ht]
  \centering
  \includegraphics[width=0.45\textwidth]{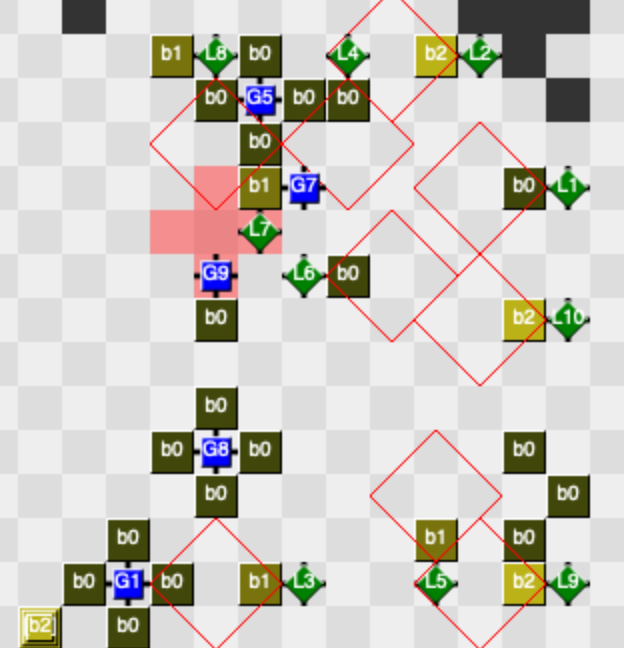}
  \caption{Step 191: \lfc clearing their blocks.}
  \label{FIG:lfc191}
\end{figure}

Naturally, this simulation might be interesting since it was the only one that \lfc lost. The final score was 130 to 40 for \dtu. If we look at the completed tasks, we see that \dtu was already able to submit a task in step 78, which yielded 90 points, since the required shape consisted of three blocks. After this however, for more than 200 steps ``nothing'' happens. The next task is completed, again by \dtu, in step 317, netting 40 points for a two-block shape. \lfc only completes one task, in step 351, receiving the 40 points. After this, no further tasks are completed.
So, one question is surely what did \lfc do before step 351. Reviewing all other simulations of \lfc, the agents were always able to complete their first task around step 200. In step 191, we find instead the situation depicted in Fig.~\ref{FIG:lfc191}.

The green \lfc agents (diamond-shaped, labels start with an ``L'', located at the top and right of this clipping), are all charging a \texttt{clear} action (red diamond markers on the grid) to remove the block they have currently attached. From their usual strategy, we conclude that their plan was to attach blocks to agent \texttt{L7}, who was already waiting in the red goal zone. If we go back in time, we see that the \dtu agents are and have been very active in this region, carrying lots of blocks, as always. First, this makes it very difficult for \lfc to get their blocks to the \texttt{L7} agent. Secondly, the \lfc agents decide to abandon their whole plan, even clearing all blocks they have already gathered. If we assume that \lfc has to start anew (minus some initial discovery and exploration), we might indeed expect the next task to be completed after another 100 to 150 steps, which proved to be the case.

\subsubsection{2nd simulation of \lfc vs. \trg} 
In this simulation both teams were not able to score. This is quite surprising since \lfc scored in all simulations but this one. Moreover, \lfc was the team that scored most in the contest: 1790 points. Considering only simulations against \trg, it scored 180 in the first, and 210 points in the third. The question is: Why did \lfc perform much better in those other simulations? 

\begin{figure}
    \centering
    \includegraphics[scale=0.6]{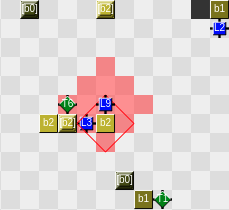}
    \caption{The exact moment (step 174) when a \trg agent disables \lfc's agents.}
    \label{fig:lfc_trg}
\end{figure}

To understand that, we need to look at \trg's strategy. They always seek to position agents in the goal zone to disable any agent that enters that place. Nevertheless, it does not always work. At some times, some goal zone receives no \trg agents. As \lfc's strategy is to always choose a single goal zone to be used, at the second simulation, both strategies have collided. Every time \lfc's agents tried to deliver a task, a \trg agent was there to disable them. An example of this event is depicted in Figure~\ref{fig:lfc_trg}. 

This strategy seems to break the \lfc team, because once their delivery agent is disabled, the whole team restarts to clear blocks and search for the grid's boundaries again. Whilst \lfc is unable to deliver tasks, so is \trg. \trg's strategy for preventing a team to score works pretty well, on the other hand, their agents were not able to coordinate to form shapes required by tasks. At the end of the second simulation, no team scored a single point.

\subsection{Survey results}

Traditionally, we conclude the \contest with a questionnaire that we ask each team to answer\footnote{The reader will find these answers at the end of each team description paper.}. At this point, we give a brief summary of the answers from all teams.

Regarding the motivation, practicing \mas development and in general learning more about agent technology were given as the main reasons for participating in the \shortmapc. This is aligned with \shortmapc's goal, in which in order to stimulate research in \mas, we need more people to learn and practice it.

We noticed that many teams mentioned that \textbf{debugging} capabilities were quite limited. Thus, they often resorted to ``print(.)'' as a debugging tool, i.e. adding logging statements and reading or searching the traces afterwards.
Some teams stated that debugging was the most time-consuming task and some even developed their own (scenario-specific) tools that helped them to understand what was going on.
Other time-consuming tasks include map navigation and merging the local views of the agents.

The most challenging aspects of the scenario according to the teams were:
\begin{itemize}
    \item the dynamic environment,
    \item the local perspective of the agents, and
    \item coordinating agents to perform the synchronous actions.
\end{itemize}

From the survey, we also know that the teams barely added additional AI techniques to improve their solution (aside from \lfc using a fast downward planner). This is probably due to the additional time investments required to add features to systems that are already quite complex within a limited time-span.

The main advantages of using agent technology were seen as flexibility and modularity of the system. From an agent programming perspective, agents should consider constantly the current state to select a proper action which may be an useful feature in dynamic environments. As the main drawbacks, teams cited the difficulty in debugging, a lack of portability and that it was very challenging to keep the system simple and easy to maintain.

Finally, if teams were to attend another time, they would like to improve error handling, reliability, coordination of their agents and their own debugging means.

\section{Lessons learned}

The organization of our \shortmapc turned out to be quite work-intensive at times.
Its technical implementation has been mainly done or supervised by (to this day seven) PhD students of the second author, in addition to a number of  Bachelor and Master theses. However, the students also played a major role in crafting the scenarios and coming up with fresh and innovative ideas.

In the first phase of the \shortmapc, no real cooperation among agents was achieved. In fact \emph{every man for himself} was a common strategy, completely against the paradigm of agent programming. Often the teams with the best working $A^*$ path-finding algorithm won. Due to the fact that the participating agent languages were not yet mature enough, the main benefit  of the contest in the early days was to serve as a debugging tool for the participating systems.

Indeed, low-level technical problems with the implementations of the agent languages often played a major role. This is in contrast to the second phase, where attention shifted to the scenario and higher-level
concerns.
 
\subsection{Agents assemble scenario}
In the new scenario, we saw that forcing agents to work solely off their local perspectives and integrate their knowledge with other agents is a challenging task. 

We once again note that it's desirable to have a problem that is easy to solve, but very difficult to solve well. In other words, it should be easy to come up with some agents that can play the game, while mastering it should require a lot of effort.

Aside from \trg trying to defend goal zones, we only saw limited conscious interaction between the teams. Unfortunately, our options to elicit interaction are also limited, because there is little motivation to cooperate in a zero-sum game. As such, it would only work if both teams are deceived to varying degrees. Another way would be some form of attacking, though we try to keep our scenarios as peaceful as possible.
In this scenario, we had indirect interaction through presence and modification of a shared environment, similar to the \emph{Cow} scenario. In the \emph{City} scenario, we had very limited interaction followed by the well-building attempt. In the \emph{Mars} scenario, we had interaction through attacking agents, though the extent (duration and complexity) of these interactions also remained expandable. A challenge for the future is surely to design complex interactions which are interesting to realize and see in action.

\subsection{General}

A lesson of the early phase was the awareness that normally neglected \emph{engineering} issues (as opposed to \emph{scientific} ones) are of utmost importance. For example, collecting statistical data or providing visualizations turned out to be as  important as the choice and the tuning of the scenarios. Without them it was extremely difficult to analyze why a team behaved as it did.
  
Using automatically generated statistical data, we can easily retrace a whole simulation's progress by looking at the generated charts instead of watching the whole replay. The charts mainly focused on scenario-specific data, like the development of the score or stability of dominated zones. Furthermore, we were finally able to directly and easily compare different simulation runs without having to keep a lot of details in mind. 
Such tools cannot only be used for debugging the teams' agents, but also for analysis of the scenario and improving it for the next round.
   
These insights went into the \emph{Agents-on-Mars} scenario, where we noted an increasing number of multi-agent platforms. Since then, our scenarios have always been won by dedicated agent platforms---they seem to outperform ``ad-hoc'' solutions. This might be attributed to some teams taking part repeatedly, but it also points to an increasing maturity and ease of use concerning multi-agent platforms. 

In order to better understand the underlying strategies of the teams, we worked out a standardized questionnaire \cite{emas13quest} (which was further improved over the years). This did not only help to learn about the systems and the results they produced, but also to understand the whole development process. Additionally, it serves for newcomers to avoid mistakes from previous iterations.

The motivation to enter the \shortmapc was for some teams simply to learn about multi-agent systems or to refine  programming skills. Furthermore, most teams shared our goal of evaluating multi-agent frameworks and platforms. Regarding their structure, teams were composed of students as well as researchers with their background mostly in \mas or at least in artificial intelligence in general. This reinforced our motivation to always come up with new scenarios, rather than optimizing a particular one over the years, which only favors teams that attend each and every year (it seems this happened in the simulation league of the soccer competition).

We also asked the teams how difficult it was and how much effort had to be put into getting to a point where their system behaved as it finally did. We got very diverse results, ranging from only a hundred to over a thousand hours and 1000 to many thousands of lines of code that had to be written, tested and debugged. This clearly hints at varying levels of usability concerning different agent platforms.

Furthermore, teams noted that they not only debugged their agents but found and fixed bugs in the agent framework or platform they used as well, which shows that the \shortmapc can play a major role concerning the development and evaluation of different platforms. Nevertheless, the teams are still not satisfied with available state-of-the-art debugging tools, since it  requires a lot of effort to debug even 20 agents, each with its own individual mindset.

Lastly, we realized that the visualization and playability of the respective scenario is a key to reaching a broader audience, especially students, e.g. when \massim is used in teaching in various courses all over the world. The competitive nature is fun for the students and this feature should never be underestimated.

\section{Challenges and outlook}

While the agent paradigm plays an important role in computer science, its uptake in industry still remains small. We believe that the \shortmapc plays some role in determining under which conditions agent languages can be used in practice.

The ideal scenario we were (and still are) looking for, should be easily testable and not be based on  difficult rules (only the  solutions should be difficult), so that beginners in the area of agent programming can easily take on the challenge. A good solution should use cooperation among autonomous agents and be flexible enough so that different groups of agents evolve and work together to solve intermediate tasks.
 
After almost 15 years of research and experience, we still have not found such a \emph{convincing scenario}. Nor have we yet proved that agent-based approaches are clearly superior to other, sometimes even ad-hoc, approaches using traditional programming languages. In many areas of computer science, one is often looking for a \emph{killer application}. However, it may well be that such killer applications do not exist.
In defense of \mas, there are many potential advantages that the contest is not evaluating at all, because it does not seem feasible in the context of the competition: Reusability, maintenance, correctness, the possibility to model-check agents, code running on different platforms, etc.

Regardless, there are good reasons to be optimistic, because there is progress on two sides. First, multi-agent programming technologies are becoming more and more capable. Secondly, there are many lessons learned throughout the history of the contest and we are getting better at encouraging the cooperative behaviors we want to see in agents.
 
So what are possible ways to improve our \shortmapc? We are considering the idea to let \emph{more than two} teams participate in the same simulation. The current scenario would provide for this naturally, however, we need to address better visualization (too many things happening at the same time) and evaluation (to easily find interesting situations and emergent behavior) first.

Our ultimate vision is an agent platform that allows to deploy agents written in very different agent languages, using the specific features of them. For example, it might be beneficial for BDI agents to solve very efficiently certain tasks, whereas planning agents based on some form of \emph{hierarchical task nets} could do the planning for them. Being able to re-use agents already developed (and based on different paradigms) would certainly push the envelope for applications of multi-agent systems in general.

Agents running on a local platform (rather than participating over the Internet) would also allow more fine-grained control over communication and real-time aspects.
We could then consider \emph{many} agents, not just a few, but hundreds or thousands of \emph{sophisticated} agents---traditional approaches  do not seem to perform well in such a situation. Moreover, with many \emph{interacting agents} we might see some interesting behavior evolve.

However, the price to pay is to standardize the communication and set up common protocols and interfaces for such agents. That would change our contest drastically.

\section*{Acknowledgment}
The authors would like to thank Alfred Hofmann from Springer for his continuous support right from the beginning, and for endowing the price of $500$ Euros in Springer books.

Last but not least, we thank all teams that entered our \contest in the last 15 years and helped to keep it alive. The second author sends a special thank you to some former PhD students in his group: Peter Novák, Tristan Behrens, Federico Schlesinger, and Michael Köster. Without their ideas and enthusiasm, the \shortmapc 
would not have flourished.

\bibliographystyle{splncs04}
\bibliography{main}

\end{document}